\documentclass[aps,prb,twocolumn,superscriptaddress,showpacs,floatfix,amsmath,amssymb]{revtex4-1}
\usepackage{graphicx,xcolor}
\usepackage{mathtools}

\def\etal{{\em{et al}}}	

\begin{document}
 
\title{Study of multiband disordered systems using the typical medium dynamical cluster approximation}
 
\author{Yi Zhang}
\email{zhangyiphys@gmail.com}
\affiliation{Department of Physics \& Astronomy, Louisiana State University, Baton Rouge, Louisiana 70803, USA}
\affiliation{Center for Computation \& Technology, Louisiana State University, Baton Rouge, Louisiana 70803, USA}
\author{Hanna Terletska}
\affiliation{Department of Physics, University of Michigan, Ann Arbor, Michigan 48109, USA}
\author{C. Moore}
\affiliation{Department of Physics \& Astronomy, Louisiana State University, Baton Rouge, Louisiana 70803, USA}
\affiliation{Center for Computation \& Technology, Louisiana State University, Baton Rouge, Louisiana 70803, USA}
\author{Chinedu Ekuma}
\affiliation{Department of Physics \& Astronomy, Louisiana State University, Baton Rouge, Louisiana 70803, USA}
\affiliation{Center for Computation \& Technology, Louisiana State University, Baton Rouge, Louisiana 70803, USA}
\author{Ka-Ming Tam}
\affiliation{Department of Physics \& Astronomy, Louisiana State University, Baton Rouge, Louisiana 70803, USA}
\affiliation{Center for Computation \& Technology, Louisiana State University, Baton Rouge, Louisiana 70803, USA}
\author{Tom Berlijn}
\affiliation{Center for Nanophase Materials Sciences, Oak Ridge National Laboratory, Oak Ridge, TN 37831, USA}
\affiliation {Computer Science and Mathematics Division, Oak Ridge National Laboratory, Oak Ridge, Tennessee 37831, USA}
\author{Wei Ku}
\affiliation{Condensed Matter Physics and Materials Science Department, Brookhaven National Laboratory, Upton, New York 11973, USA}
\affiliation{Physics Department, State University of New York, Stony Brook, New York 11790, USA}
\author{Juana Moreno}
\affiliation{Department of Physics \& Astronomy, Louisiana State University, Baton Rouge, Louisiana 70803, USA}
\affiliation{Center for Computation \& Technology, Louisiana State University, Baton Rouge, Louisiana 70803, USA}
\author{Mark Jarrell}
\affiliation{Department of Physics \& Astronomy, Louisiana State University, Baton Rouge, Louisiana 70803, USA}
\affiliation{Center for Computation \& Technology, Louisiana State University, Baton Rouge, Louisiana 70803, USA}

\begin{abstract}
We generalize the typical medium dynamical cluster approximation to
multiband disordered systems. Using our extended formalism, we perform 
a systematic study of the non-local correlation effects induced by disorder on 
the density of states and the mobility edge of the three-dimensional two-band 
Anderson model. We include inter-band and intra-band hopping and an intra-band 
disorder potential.  Our results are consistent with the ones obtained by the 
transfer matrix and the kernel polynomial methods. We apply the method to
K$_x$Fe$_{2-y}$Se$_2$ with Fe vacancies. Despite the strong vacancy disorder
and anisotropy, we find the material is not an Anderson insulator.  
Our results demonstrate the application of the typical medium dynamical cluster approximation method 
to study Anderson localization in real materials. 
\end{abstract}

\pacs{71.23.An,72.80.Ng,71.10.Fd,74.70.-b}

\maketitle
 
\section{Introduction}
\label{sec:intro} 
The role of disorder (randomness) in materials has been at the forefront of current 
research~\cite{Lee_RevModPhys,Belitz_RevModPhys,50years} due 
to the new and improved functionalities that can be achieved in materials by 
carefully controlling the concentration of impurities in the host. At half-filling and in the absence of any 
spontaneous symmetry breaking field, disorder can induce a transition in a non-degenerate
electronic three-dimensional system from a metal to an insulator (MIT)~\cite{Anderson,Kramer}. 
This phenomenon, which occurs due to the multiple scattering of charge carriers off random impurities, 
is known as Anderson localization~\cite{Anderson}.

The most commonly used mean-field theory to study disordered systems is the coherent potential approximation 
(CPA)~\cite{PhysRev.156.809,Velicky68,Kirpatrick-Velickyprb70}, which maps the original disordered 
lattice to an impurity embedded in an effective medium. The CPA successfully describes some one-particle properties, 
such as the average density of states (ADOS) in substitutional disordered alloys~\cite{PhysRev.156.809,Velicky68,Kirpatrick-Velickyprb70}. 
However, being a single-site approximation, 
the CPA by construction neglects all disorder-induced nonlocal correlations involving multiple scattering processes.
To remedy this, cluster extensions of the CPA such as the dynamical cluster approximation 
(DCA)~\cite{PhysRevB.61.12739,PhysRevB.63.125102,Jarrell01} and the molecular CPA~\cite{MCPA}
have been developed, where nonlocal effects are incorporated. Unfortunately,
all of these methods fail to capture the Anderson localization transition
since the ADOS utilized in these approaches is neither critical at the transition or distinguish 
the extended and the localized states. 

In order to describe the Anderson transition in such effective medium theories, a proper order 
parameter has to be used. As noted by Anderson, the probability distribution of the local density of states (LDOS) must be considered, 
and the most probable or typical value would characterize it~\cite{Anderson,RevModPhys.50.191}. It was found that 
the geometric mean of the LDOS is a good approximation of its typical value 
(TDOS) and it is critical at the transition~\cite{Janssen1998,Vollhardt,Crow1988}, 
which makes it an appropriate order parameter to describe Anderson localization. Based on this idea, 
Dobrosavljevic \etal.~\cite{Vlad2003} formulated a single-site typical medium theory 
(TMT) for Anderson localization which gives a qualitative description of the transition in three 
dimensions. In contrast to the CPA, the TMT uses the geometrical averaging over the disorder configuration 
in the self consistency loop. And thus, the typical not the average DOS is used as the order parameter.
However, due to the single-site nature of the TMT it neglects nonlocal correlations such 
as the effect of coherent back scattering.  Thus, the TMT underestimates the critical disorder strength of the
Anderson localization transition and fails to capture the reentrant 
behavior of the mobility edge (which separates the extended and localized states) for uniform box disorder. 

Recently, a cluster extension of TMT was developed, named the typical medium dynamical cluster approximation 
(TMDCA)~\cite{PhysRevB.89.081107},  which predicts accurate critical disorder strengths and captures the reentrant 
behavior of the mobility edge. The TMDCA was also extended to include off-diagonal in addition to 
diagonal disorder.~\cite{PhysRevB.90.094208}.  However, like the TMT, the previous TMDCA implementations
have only been developed for single-band systems, and in
real materials, there are usually more than one band close to the Fermi level. 
Sen performed CPA calculation on two-band semiconducting binary alloys~\cite{PhysRevB.8.5613}, and the 
electronic structure of disordered systems with multiple bands has also been studied numerically in finite 
systems~\cite{Aoki1,Aoki2}. But a good effective medium theory to study Anderson localization transition 
in multiband systems is still needed to understand the localization phenomenon in real systems such as 
diluted doped semi-conductors, disordered systems with strong spin-orbital coupling, etc.

In this paper, we extend the TMDCA to multiple band disordered systems with both intra-band and inter-band 
hopping, and study the effect of intra-band disorder potential on electron localization. We perform 
calculations for 
both single-site and finite size clusters, and compare the results with those from numerically exact 
methods, including transfer matrix method (TMM) and kernel polynomial method (KPM).  We show that finite 
sized clusters are necessary to include the nonlocal effects and produce more accurate results. Since these
results show that the method is accurate and systematic, we then apply it to study the iron selenide 
superconductor K$_x$Fe$_{2-y}$Se$_2$ with Fe vacancies, as an example to show that this method can be used 
to study localization effects in real materials.  In addition, as an effective medium theory, our method is also 
able to treat interactions~\cite{2015arXiv150300025E}, unlike the TMM and KPM.

The paper is organized as follows. We present the model and describe the details of the formalism 
in Sec.~\ref{sec:formalism}. In Sec.~\ref{sec:results2band},  we present our results of the ADOS and 
TDOS for a two-band disordered system with various parameters, and use the vanishing of the TDOS 
to: determine the critical disorder strength, extract the mobility edge and construct a 
complete phase diagram in the disorder-energy parameter space for different inter-band hopping.
In Sec.~\ref{sec:KFeSe}, we discuss simulations of K$_x$Fe$_{2-y}$Se$_2$ with Fe vacancies.
We summarize and discuss future directions in Sec.~\ref{sec:conclusion}.  In Appendix \ref{appendixA},
we provide justification for the use of our order parameter ansatz.

\section{Formalism}
\label{sec:formalism}
\subsection{Dynamical cluster approximation for multiband disordered systems}
We consider the multiband Anderson model of non-interacting electrons with nearest neighbor hopping and 
random on-site potentials. 
The Hamiltonian is given by
\begin{equation} \label{eqn:model}
\begin{split}
H=&-\sum_{<ij>} \sum_{\alpha,\beta=1}^{l_b}t_{ij}^{\alpha\beta} (c_{i\alpha}^{\dagger} c_{j\beta} + c_{j\beta}^{\dagger} c_{i\alpha})\\
&+\sum_{i=1}^N \sum_{\alpha,\beta=1}^{l_b} (V_{i}^{\alpha\beta} - \mu \delta_{\alpha\beta}) n_i^{\alpha \beta}
\end{split}
\end{equation}
The first term provides a realistic multiband description of the host valence bands. The labels
$i$, $j$ are site indices and $\alpha, \beta$ are band indices. The operators
 $c_{i\alpha}^\dagger$($c_{i\alpha}$) 
create (annihilate) a quasiparticle on site $i$ and band $\alpha$. The second part denotes the disorder, which is 
modeled by a local potential $V_i^{\alpha \beta}$ that is randomly distributed according to some specified probability 
distribution $P(V_i^{\alpha \beta})$, where $n_{i}^{\alpha\beta} = c_{i\alpha}^\dagger c_{i\beta}$, 
$\mu$ is the chemical potential,  and $t_{ij}^{\alpha\beta}$ are the hopping matrix elements. 
%We set 4$t$ = 1 as the energy unit.  
Here we consider binary disorder, where the random on-site 
potentials $V_i^{\alpha\beta}$ obey independent binary probability distribution 
functions with the form
\begin{equation}
P(V_i^{\alpha\beta})=x\delta(V_i^{\alpha\beta}-V_A^{\alpha\beta})+(1-x)\delta(V_i^{\alpha\beta}-V_B^{\alpha\beta}).
\end{equation}

In our model, there are $l_b$ band indices so that both the hopping and disorder potential 
are $l_b \times l_b$ matrices. The random potential is %(the underbar denotes $l_b \times l_b$ matrices in the band basis)
\begin{equation}
\underline{V_i} = \\
\left(\begin{array}{cccc}
V_{i}^{\alpha\alpha}    	& \cdots	 & V_{i}^{\alpha\beta} 	 \\
        .		 	&	 .	 &     .		 \\
	.		 	&	 .	 &     .		 \\
	.		 	&	 .	 &     .		 \\
V_{i}^{\alpha\alpha}		 	& \cdots	 &  V_{i}^{\beta\alpha} \\
\end{array}\right),
\label{eqn:potential_MB}
\end{equation}
while the hopping matrix is
\begin{equation}\label{eqn:hopping}
\underline{t_{ij}} = \\
\left(\begin{array}{cccc}
t_{ij}^{\alpha\alpha}    	& \cdots	 & t_{ij}^{\alpha\beta} 	 \\
        .		 	&	 .	 &     .		 \\
	.		 	&	 .	 &     .		 \\
	.		 	&	 .	 &     .		 \\
t_{ij}^{\beta\alpha}		 	& \cdots	 &  t_{ij}^{\beta\beta} \\
\end{array}\right),
\end{equation}
%\begin{equation}
%\underline{t_{ij}}=\underline{t}=\left(\begin{array}{cc}
%t^{aa} & t^{ab}\\
%t^{ba} & t^{bb}
%\end{array}\right)
%\end{equation}
%\begin{equation}
%\underline{V_i}=\left(\begin{array}{cc}
%V_i^{aa} & V_i^{ab}\\
%V_i^{ba} & V_i^{bb}
%\end{array}\right)
%\end{equation}
where underbar denotes $l_b \times l_b$ matrix, $t^{\alpha \alpha}$ and $t^{\beta \beta}$ are intra-band hoppings, 
while $t^{\alpha \beta}$ and $t^{\beta \alpha}$ are inter-band hoppings. Similar definitions apply 
to the disorder potentials.
If we restrict the matrix elements to be real, Hermiticity requires both matrices to be symmetric, i.e., $t^{\alpha \beta}=t^{\beta \alpha}$ and 
$V_i^{\alpha \beta}=V_i^{\beta \alpha}$.  

To solve the Hamiltonian of Eq.~\ref{eqn:model}, we first generalize the standard DCA to a multiband system.
Within DCA the original lattice model is mapped onto a
cluster of size $N_c=L^3$ with periodic boundary condition embedded in an effective medium. 
The first Brillouin zone is divided in $N_c$ coarse 
grained cells~\cite{PhysRevB.63.125102},
whose center is labeled by $K$, surrounded by points labeled by $\tilde{k}$ within the cell.
Therefore, all the $k$-points are expressed as $k=K+\tilde{k}$. The
effective medium is characterized by the hybridization function $\underline{\Delta(K,\omega)}$. 
The generalization of the DCA to a multiband system entails representing all the quantities in momentum 
space as $l_b \times l_b$ matrices.

The DCA self-consistency loop starts with an initial guess for the hybridization matrix $\underline{\Delta(K,\omega)}$,
which is given by
\begin{equation}
\underline{\Delta(K,\omega)} = \\
\left(\begin{array}{cccc}
\Delta^{\alpha\alpha}(K,\omega)    	& \cdots	 & \Delta^{\alpha\beta}(K,\omega) 	 \\
        .		 	&	 .	 &     .		 \\
	.		 	&	 .	 &     .		 \\
	.		 	&	 .	 &     .		 \\
\Delta^{\beta\alpha}(K,\omega)		 	& \cdots	 &  \Delta^{\beta\beta}(K,\omega) \\
\end{array}\right).
\label{eq:gamma_MB}
\end{equation}

%\begin{equation}
%\underline{\Delta(K,\omega)}= \left(\begin{array}{cc}
%\Delta^{aa}(K,\omega) & \Delta^{ab}(K,\omega)\\
%\Delta^{ba}(K,\omega) & \Delta^{bb}(K,\omega)
%\end{array}\right).
%\end{equation}
For the disordered system, we must solve the cluster problem in real space.
In that regard, for each disorder configuration described by the disorder potential $V$ we calculate
the corresponding cluster Green function which is now an $l_b N_c \times l_b N_c$ matrix
\begin{equation}
\underline{G}_{c}(V)= \left({\omega}\mathbb{I}-{\underline{t}^{(\alpha\beta)}}-
{\underline{\Delta'}^{(\alpha\beta)}}-{\underline{V}^{\alpha\beta}}\right)^{-1} 
\label{eq:cgreen}\,.
\end{equation}
Here, $\mathbb{I}$ is identity matrix and $\Delta_{ij}^{'}$ is the Fourier transform (FT) of the hybridization, 
i.e.,
%\begin{equation}
% G_c(V)=(\omega I - \bar{t}' -\Delta' -V)^{-1}
%\end{equation}
%where V is diagonal in both real space and band space as we only consider the 
%intra-band disorder, and $\bar{t}'$ and $\Delta'$ are calculated by Fourier
%transforming the hopping and hybridization matrices to real space as
%\begin{equation}
% t_{ij}^{'\alpha\beta} = \sum_K \epsilon^{\alpha\beta}(K) exp[iK\cdot(r_i -r_j)]
%\end{equation}
\begin{equation}
 \Delta_{ij}^{'\alpha\beta} = \sum_K \Delta^{\alpha\beta}(K) exp[iK\cdot(r_i -r_j)].
\end{equation}
%where $\underline{\epsilon}^{\alpha\beta}(K)$ is the matrix element of the coarse 
%grained bare dispersion $\underline{\bar{\epsilon}(K)}$ in band space,
%\begin{equation}
% \underline{\bar{\epsilon}(K)} = \frac{N_c}{N} \sum_{\tilde{k}} \underline{\varepsilon_{K+\tilde{k}}}
%\end{equation}
%so both $\bar{t}'$ and $\Delta'$ are $2N_c \times 2N_c$ matrices in real space.
 
We then stochastically sample random configurations of the disorder potential $V$ and 
average over disorder $\left\langle (\cdots)\right\rangle$ to 
get the $l_b N_c \times l_b N_c$ disorder averaged cluster Green function in real space
\begin{equation}
\underline{G_{c}(\omega)_{ij}}= \\
\left(\begin{array}{cccc}
\left\langle G_{c}^{\alpha\alpha}(\omega,V) \right\rangle_{ij}    	& \cdots	 & \left\langle G_{c}^{\alpha\beta}(\omega,V) \right\rangle_{ij} 	 \\
        .		 	&	 .	 &     .		 \\
	.		 	&	 .	 &     .		 \\
	.		 	&	 .	 &     .		 \\
\left\langle G_{c}^{\beta\alpha}(\omega,V) \right\rangle_{ij} 		 	& \cdots	 &  \left\langle G_{c}^{\beta\beta}(\omega,V) \right\rangle_{ij} \\
\end{array}\right)\,.
\label{eqn:cgreen_average}
\end{equation}
%\begin{equation}
% \underline{G_c}=\langle \underline{G_c(V)} \rangle.
%\end{equation}
We then Fourier transform to $K$ space and also impose translational symmetry to construct the $K$-dependent 
disorder averaged cluster Green function
$\underline{G_c(K,\omega)}$, which is a
$l_b \times l_b$ matrix for each $K$ component
\begin{equation}
\underline{G_c(K,\omega)}= \\
\left(\begin{array}{cccc}
G_c^{\alpha\alpha}(K,\omega)    	& \cdots	 & G_c^{\alpha\beta}(K,\omega) 	 \\
        .		 	&	 .	 &     .		 \\
	.		 	&	 .	 &     .		 \\
	.		 	&	 .	 &     .		 \\
G_c^{\beta\alpha}(K,\omega) 		 	& \cdots	 &  G_c^{\beta\beta}(K,\omega) \\
\end{array}\right) .
\label{eq:KspaceG}
\end{equation}
%\begin{equation}
% \underline{G_c(K,\omega)}=\left(\begin{array}{cc}
%G_c^{aa}(K,\omega) & G_c^{ab}(K,\omega)\\
%G_c^{ba}(K,\omega) & G_c^{bb}(K,\omega)
%\end{array}\right).
%\end{equation}
After the cluster problem is solved, we can calculate the coarse grained lattice Green function matrix
%\begin{widetext}
\begin{eqnarray}\label{coarsegraining_MB}
\underline{\overline{G}(K,\omega)} &=& 
\left(\begin{array}{cccc}
\overline{G}^{\alpha\alpha}(K,\omega)    	& \cdots	 & \overline{G}^{\alpha\beta}(K,\omega) 	 \\
        .		 	&	 .	 &     .		 \\
	.		 	&	 .	 &     .		 \\
	.		 	&	 .	 &     .		 \\
\overline{G}^{\beta\alpha}(K,\omega) 		 	& \cdots	 &  \overline{G}^{\beta\beta}(K,\omega) \\
\end{array}\right) \\ \nonumber
 &=&  \frac{N_{c}}{N}\sum_{\tilde{k}}\Big(\underline{ G_{c}(K,\omega)}^{-1}+\underline{\Delta{}(K,\omega)} - \underline{\varepsilon_{k}}+\underline{\overline{\epsilon}(K)}\Big)^{-1},
\end{eqnarray}
%\end{widetext}
where the overbar denotes cluster coarse-graining, and $\underline{\overline{\epsilon}(K)}$ is the 
cluster coarse-graining Fourier transform of the kinetic energy 
\begin{equation}
%\label{eqn:coarse grained}
\underline{\overline{\epsilon}(K)} = \underline{E_0} + \frac{N_{c}}{N}\sum_{\tilde{k}} \underline{\varepsilon_{k}}
%\underline{\overline{\epsilon}(K)} = \\
%\left(\begin{array}{cccc}
%E_0^{\alpha\alpha}+\frac{N_{c}}{N}\sum_{\tilde{k}} \varepsilon_{k}    	& \cdots	 & E_0^{\alpha\beta}+\frac{N_{c}}{N}\sum_{\tilde{k}} \varepsilon_{k} 	 \\
%        .		 	&	 .	 &     .		 \\
%	.		 	&	 .	 &     .		 \\
%	.		 	&	 .	 &     .		 \\
%E_0^{\beta\alpha}+\frac{N_{c}}{N}\sum_{\tilde{k}} \varepsilon_{k}		 	& \cdots	 &  E_0^{\beta\beta}+\frac{N_{c}}{N}\sum_{\tilde{k}} \varepsilon_{k} \\
%\end{array}\right)
\end{equation}
where $E_0^{\alpha\beta}$ is a local energy, which is used to shift the bands.
The diagonal 
components of Eq.~\ref{coarsegraining_MB} have the same normalization than a conventional, i.e., scalar, Green function.
%\begin{equation}
%\begin{split}
% \underline{\bar{G}(K,\omega)}&=\left(\begin{array}{cc}
%\bar{G}^{aa}(K,\omega) & \bar{G}^{ab}(K,\omega)\\
%\bar{G}^{ba}(K,\omega) & \bar{G}^{bb}(K,\omega)
%\end{array}\right)
\\
%&=\frac{N_c}{N} \sum_{\tilde{k}} \underline{G_c(K,\omega)}^{-1} + \underline{\Delta(K,\omega)} -\underline{\varepsilon_{\tilde{k}}} + \underline{\bar{\epsilon}(K)})^{-1}.
%\end{split}
%\end{equation}

The DCA self-consistency condition requires the disorder averaged cluster Green function equal 
the coarse grained lattice Green function
\begin{equation}
 \underline{G_c(K,\omega)} = \underline{\bar{G}(K,\omega)}.
\end{equation}

Then, we close our self-consistency loop by updating the hybridization function matrix using linear mixing
\begin{equation}\label{eqn:update}
 \underline{\Delta_n(K,\omega)} = \underline{\Delta_o(K,\omega)} + \xi [\underline{G_c^{-1}(K,\omega)} - \underline{\bar{G}^{-1}(K,\omega)}], 
\end{equation}
where the subscript $``o"$ and $``n"$ denote old and new respectively, and $\xi$ is a linear mixing factor $0<\xi<1$. 
The procedure above is repeated until
the hybridization function matrix converges to the desirable accuracy $\underline{\Delta_n(K,\omega)} = \underline{\Delta_o(K,\omega)}$.

We can see that when the inter-band hopping, $t^{\alpha\beta}$, and disorder potential, $V^{\alpha\beta}$, 
vanish all the $l_b \times l_b$ matrices become diagonal, and the formalism reduces to single band DCA for 
$l_b$ independent bands.

\subsection{Typical medium theory for multiband disordered systems}
To study localization in multiband systems, we generalize the recently developed TMDCA~\cite{PhysRevB.89.081107} 
where the TDOS is used as the order parameter of the Anderson localization transition, so 
the electron localization is captured by the vanishing of the TDOS. We will use this 
TMDCA formalism to address the question of localization and mobility edge evolution
in the multiband model.

Unlike the standard DCA, where the Green function is averaged over disorder algebraically, the TMDCA 
calculates the typical (geometrically) averaged cluster density of states in the self-consistency loop as
\begin{equation}
 \rho_c^{typ}(K,\omega) = e^{\frac{1}{Nc}\sum_{i}\left\langle \log \rho_{ii}(\omega)\right\rangle}\left\langle\frac{\rho(K,\omega)}{\frac{1}{Nc}\sum_{i}\rho_{ii}(\omega)}\right\rangle,
\end{equation}
which is constructed as a product of the geometric average of the local density of 
states, $\rho_{ii}=-\frac{1}{\pi}Im G_{ii}(\omega)$, and the linear
average of the normalized momentum resolved density of states $\rho(K,\omega)=-\frac{1}{\pi}Im G_c(K,\omega)$.
The cluster-averaged typical Green function
is constructed via the Hilbert transformation
\begin{equation}
 G_c^{typ}(K,\omega)=\int d\omega' \frac{\rho_c^{typ}(K,\omega')}{\omega-\omega'}.
\end{equation}
Generalization of the TMDCA to the multiband case is not straightforward since the 
off-diagonal LDOS $\rho_{ii}^{\alpha\beta}(\omega)=-\frac{1}{\pi}G_{ii}^{\alpha\beta}(\omega)$
is not positive definite. We construct the $l_b \times l_b$ matrix for the typical density of states as
\begin{widetext}
\begin{equation}
\underline{\rho^c_{typ}(K,\omega)}= \\
\left(\begin{array}{cccc}
e^{\frac{1}{Nc}\sum_{i}\left\langle ln\rho_{ii}^{\alpha\alpha}(\omega)\right\rangle}\left\langle\frac{\rho^{\alpha\alpha}(K,\omega)}{\frac{1}{Nc}\sum_{i}\rho_{ii}^{\alpha\alpha}(\omega)}\right\rangle    	& \cdots	 & e^{\frac{1}{Nc}\sum_{i}\left\langle ln|\rho_{ii}^{\alpha\beta}(\omega)|\right\rangle}\left\langle\frac{\rho^{\alpha\beta}(K,\omega)}{\frac{1}{Nc}\sum_{i}|\rho_{ii}^{\alpha\beta}(\omega)|}\right\rangle 	 \\
        .		 	&	 .	 &     .		 \\
	.		 	&	 .	 &     .		 \\
	.		 	&	 .	 &     .		 \\
e^{\frac{1}{Nc}\sum_{i}\left\langle ln|\rho_{ii}^{\beta\alpha}(\omega)|\right\rangle}\left\langle\frac{\rho^{\beta\alpha}(K,\omega)}{\frac{1}{Nc}\sum_{i}|\rho_{ii}^{\beta\alpha}(\omega)|}\right\rangle 		 	& \cdots	 &  e^{\frac{1}{Nc}\sum_{i}\left\langle ln\rho_{ii}^{\beta\beta}(\omega)\right\rangle}\left\langle\frac{\rho^{\beta\beta}(K,\omega)}{\frac{1}{Nc}\sum_{i}\rho_{ii}^{\beta\beta}(\omega)}\right\rangle  \\
\end{array}\right).
\label{eqn:ansatz}
\end{equation}
%\begin{equation}\label{eqn:ansatz}
%\underline{\rho_{typ}(K,\omega)}=\left(\begin{array}{cc}
%e^{\frac{1}{Nc}\sum_{i}\left\langle ln\rho_{ii}^{aa}(\omega)\right\rangle}\left\langle\frac{\rho^{aa}(K,\omega)}{\frac{1}{Nc}\sum_{i}\rho_{ii}^{aa}(\omega)}\right\rangle & e^{\frac{1}{Nc}\sum_{i}\left\langle ln|\rho_{ii}^{ab}(\omega)|\right\rangle}\left\langle\frac{\rho^{ab}(K,\omega)}{\frac{1}{Nc}\sum_{i}|\rho_{ii}^{ab}(\omega)|}\right\rangle\\
%e^{\frac{1}{Nc}\sum_{i}\left\langle ln|\rho_{ii}^{ba}(\omega)|\right\rangle}\left\langle\frac{\rho^{ba}(K,\omega)}{\frac{1}{Nc}\sum_{i}|\rho_{ii}^{ba}(\omega)|}\right\rangle & e^{\frac{1}{Nc}\sum_{i}\left\langle ln\rho_{ii}^{bb}(\omega)\right\rangle}\left\langle\frac{\rho^{bb}(K,\omega)}{\frac{1}{Nc}\sum_{i}\rho_{ii}^{bb}(\omega)}\right\rangle
%\end{array}\right).
%\end{equation}
\end{widetext}
The diagonal part takes the same form as the single-band TMDCA ansatz, and the off-diagonal 
part takes a similar form but involves the absolute value of the off-diagonal `local' density of states.

We construct the typical cluster Green function through a Hilbert transformation
\begin{equation}
\underline{ G_{typ}^c(K,\omega)}= \\
\left(\begin{array}{cccc}
\int d\omega'\frac{\rho_{typ}^{\alpha\alpha}(K,\omega')}{\omega-\omega'}    	& \cdots	 & \int d\omega'\frac{\rho_{typ}^{\alpha \beta}(K,\omega')}{\omega-\omega'} 	 \\
        .		 	&	 .	 &     .		 \\
	.		 	&	 .	 &     .		 \\
	.		 	&	 .	 &     .		 \\
\int d\omega'\frac{\rho_{typ}^{\beta \alpha}(K,\omega')}{\omega-\omega'} 		 	& \cdots	 &  \int d\omega'\frac{\rho_{typ}^{\beta\beta}(K,\omega')}{\omega-\omega'} \\
\end{array}\right),
\label{eqn:Gtyp}
\end{equation}
%\begin{equation}\label{eqn:Gtyp}
% \underline{G_{typ}(K,\omega)} = \int d\omega' \frac{\underline{\rho_{typ}(K,\omega)}}{\omega-\omega'}
%\end{equation}
which plays the same role as $\underline{G_c(K,\omega)}$ in the DCA loop. Once $\underline{G_{typ}}$ is calculated from Eq.~\ref{eqn:Gtyp}, the 
self-consistency steps are the same as those in the multiband DCA described in the previous 
section: we calculate the coarse grained lattice 
Green function using Eq.~\ref{coarsegraining_MB}, and use it to update 
the hybridization function matrix of the effective medium via Eq.~\ref{eqn:update}.

The proposed ansatz Eq.~\ref{eqn:ansatz} has the following properties.  When the inter-band hopping 
$t^{\alpha \beta}$ and disorder potential $V^{\alpha \beta}$ vanish, it reduces to single-band TMDCA 
for $l_b$ independent bands, since all the off-diagonal elements of the
Green functions vanish. When disorder is weak, all the $V^{\alpha\alpha}$ are small so the distribution
of the LDOS becomes Gaussian with equal linear and geometric average so it reduces to DCA for a multiband disordered system.

When convergence is achieved, we use the total TDOS $\rho_{typ}^{tot}(\omega)$ to determine 
the mobility edge which is calculated
as the trace of the local TDOS matrix
\begin{equation} \label{eqn:op}
 \rho_{typ}^{tot}(\omega) = Tr\left[\frac{1}{N_c}\sum_K \underline{\rho_{typ}(K,\omega)}\right] = \sum_{ \forall \alpha =\beta} \rho_{typ}^{\alpha\beta}(\omega).
\end{equation}
This construction of the order parameter may not seem very physical as the typical value of the 
LDOS should serve as the order parameter~\cite{Anderson,RevModPhys.50.191}, and the LDOS for the multiband system
is the sum of the $l_b$ bands in the local site basis 
$\rho_i^{tot}=\sum_{\alpha =\beta} \rho_{i}^{\alpha\beta}(\omega)$.  Therefore, the real order parameter should be the typical value 
of $\rho_i^{tot}$ defined as the geometric average of the total LDOS, $\exp(\frac{1}{N_c}\sum_i \log \rho_i^{tot} )$ 
which is invariant under local unitary transformations and is not equal to the $\rho_{typ}^{tot}$ defined in Eq.~\ref{eqn:op}. 

However, Eq.~\ref{eqn:op} should also be a correct order parameter
as long as it vanishes simultaneously with the typical value of $\rho_i^{tot}$,
and we show this in Appendix~\ref{appendixA}.  By considering the distribution of the
LDOS in each band, Appendix~\ref{appendixA} shows that when localized 
states mix with extended states the system is still extended, which is consistent with
Mott's insight about the mobility edge~\cite{Mott}.  Intuitively, this makes sense as when
all the distributions of $\rho_i^{\alpha\alpha}$ are critical then the typical
values must behave as $~|V-V_c|^{\beta_\nu}$ near the transition, and so their sum must as well.
If one is not critical (on the metallic side) then Eq.~\ref{eqn:op} will not vanish as $~|V-V_c|^{\beta_\nu}$,
as expected. 

To test our multiband typical medium dynamical cluster approximation formulation, we apply it 
to the specific case of a two band model, unless otherwise stated in Sec.~\ref{sec:results}.  Throughout the discussion of our results below, we denote $\alpha$ as $a$ and $\beta$ as $b$.

\section{Results}\label{sec:results}

\subsection{Two band model}
\label{sec:results2band}
As a specific example, we test the generalized DCA and TMDCA algorithms for a three-dimensional system 
with two degenerate bands (ab) described by Eq.~\ref{eqn:model}. In this case, both the hopping and disorder potential are
2 $\times$ 2 matrices in the band basis given by
\begin{equation}
\underline{t_{ij}}=\underline{t}=\left(\begin{array}{cc}
t^{aa} & t^{ab}\\
t^{ba} & t^{bb}
\end{array}\right),
\end{equation}
and
\begin{equation}
\underline{V_i}=\left(\begin{array}{cc}
V_i^{aa} & V_i^{ab}\\
V_i^{ba} & V_i^{bb}
\end{array}\right),
\end{equation}
respectively.  The intra-band hopping is set as $t^{aa}=t^{bb}=1$, with
finite inter-band hopping $t^{ab}$.
% and a binary disorder distribution $\underline{V_A}=-\underline{V_B}=\underline{V}$.
Here, the hopping matrix is defined as dimensionless so that the bare dispersion can be written as 
$\underline{\varepsilon_k}= \underline{t} \varepsilon_k$ with $\varepsilon_k=-2t[\cos(k_x)+ \cos(k_y)+ \cos(k_z)]$ in 
three dimensions.  We choose $4t=1$ to set the units of energy.
We consider the two bands orthogonal to each other, where the local inter-band disorder $V_i^{\alpha \beta}$ vanishes and 
the randomness comes from the local intra-band disorder potential $V_i^{\alpha \alpha}$ that follow independent binary 
probability distribution functions with equal strength,  $V^{aa}=V^{bb}$. Since the two bands are degenerate and the disorder strength for each 
band is also identical, 
the calculated average DOS will be the same for each band, so we only plot the quantities for one of the 
bands in the following results, as it is enough to characterize the properties of the system.

In our formalism, in order to disorder average instead of performing the very expensive 
enumeration of all disorder configurations, which scales as $2^{2N_c}$,
we perform a stochastic sampling of configurations which greatly reduces the 
computational cost~\cite{PhysRevB.92.014209}.  This is so we can study larger systems. For a typical 
$N_c=64$ calculation, 500 disorder configurations are enough to produce reliable results 
and this number decreases with increasing cluster size.

\begin{figure}[h!]
 \includegraphics[trim = 0mm 0mm 0mm 0mm,width=1\columnwidth,clip=true]{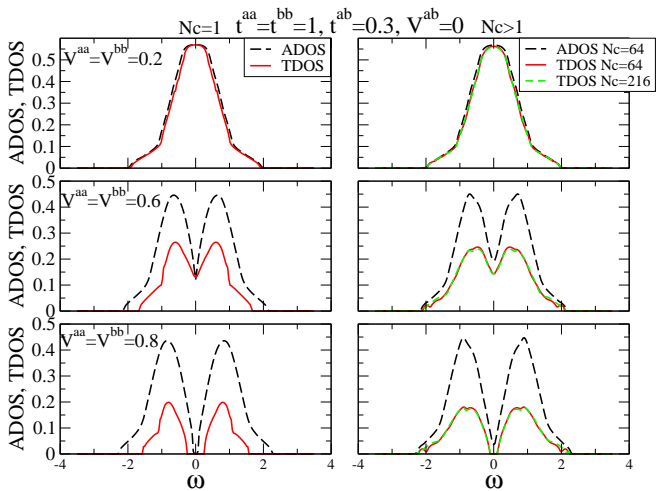}
 \caption{Evolution of the ADOS and TDOS at different disorder strengths $V^{aa}(V^{bb})$, for $N_c=1$ 
 (left panel) and $N_c>1$ (right panel) for fixed $t^{ab}=0.3$. For small disorder, the ADOS and TDOS 
 are almost identical. The TDOS is suppressed as the disorder increases. The extended states region with finite 
 TDOS for $N_c=1$ is narrower than the results of $N_c>1$ which indicates that the single-site TMT 
 overemphasizes localization.}
 \label{fig:dos_tdos_Nc}
\end{figure}

We first compare the ADOS and TDOS at various disorder strengths $V^{aa}(V^{bb})$, with a fixed inter-band 
hopping $t^{ab}=0.3$, for different cluster sizes $N_c$ in Fig.~\ref{fig:dos_tdos_Nc}. Our TMDCA scheme for 
$N_c=1$ corresponds to the analog of the TMT for two-band systems, and the ADOS is calculated with the two-band 
DCA. To show the effects of non-local correlations introduced by finite clusters, we present data for both 
$N_c=1$ and $N_c>1$. We can clearly see that the TDOS, which can be viewed as the the order parameter of 
the Anderson localization transition, gets suppressed as the disorder increases . By comparing the width 
of the extended state region, where the TDOS is finite, we can see that single site TMT overestimates  
localization.

\begin{figure}[h!]
 \includegraphics[trim = 0mm 0mm 0mm 0mm,width=1\columnwidth,clip=true]{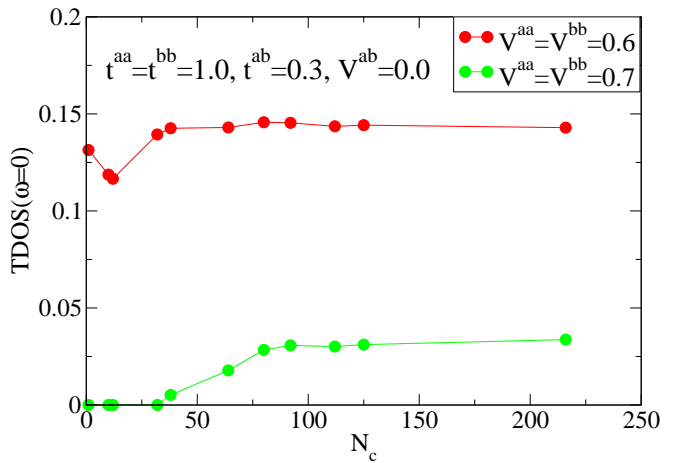}
 \caption{Evolution of the TDOS at the band center ($\omega=0$) with increasing cluster size for two different 
 sets of parameters with $t^{aa}=t^{bb}=1.0$, $t^{ab}=0.3$, $V^{ab}=0.0$, $V^{aa}=V^{bb}=0.6,0.7$. The former 
 has faster convergence (around $N_c=38$) than the latter (around $N_c=98$), due to the critical slowing 
 down closer to the transition region.}
 \label{fig:converge_Nc}
\end{figure}

\begin{figure}[h!]
 \includegraphics[trim = 0mm 0mm 0mm 0mm,width=1\columnwidth,clip=true]{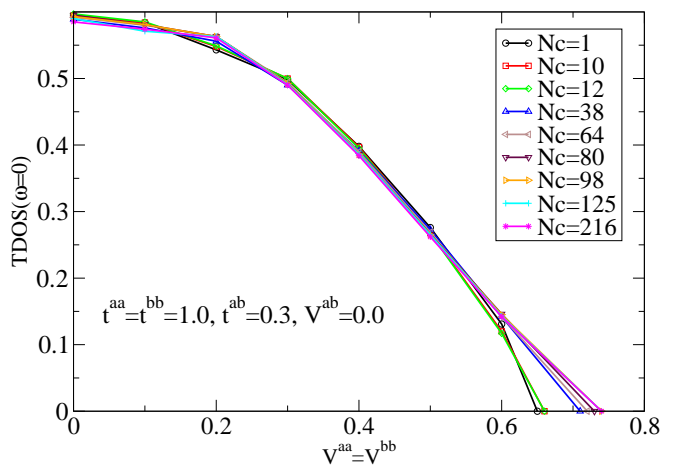}
 \caption{The TDOS at the band center ($\omega=0$) vs. $V^{aa}=V^{bb}$ with increasing cluster 
 size, for $t^{aa}=t^{bb}=1.0$, $t^{ab}=0.3$, $V^{ab}=0.0$. For $N_c=1$, the critical disorder 
 strength is 0.65 and as $N_c$ increases, it increases and converges to 0.74 for $N_c=98$.}
 \label{fig:convergence}
\end{figure}

From Fig.~\ref{fig:dos_tdos_Nc}, we see that the results of TMDCA for $N_c=64$ and $N_c=216$ 
are almost on top of each other, showing a quick convergence with the increase of cluster size. To see this 
more clearly, we plot in Fig.~\ref{fig:converge_Nc} the TDOS at the band center for two different disorder 
strengths and various cluster sizes.  We see that the results for both cases converge quickly with cluster 
size. Faster convergence (around $N_c=38$) is reached  for the case further away from the critical 
region ($V^{aa}=V^{bb}=0.6$) than for the one closer ($V^{aa}=V^{bb}=0.7$) where convergence 
is reached around $N_c=98$.  This is expected due to the critical slowing down close to the transition. 
To further study the convergence, we also plot in Fig.~\ref{fig:convergence} the TDOS at the band center as 
a function of disorder strength ($V^{aa}=V^{bb}$) for several $N_c$.  The critical disorder strength is 
defined by the vanishing of the TDOS($\omega=0$). The results show a systematic increase of the critical
disorder strength as $N_c$ increases, and the convergence is reached at $N_c=98$ with the critical value of 0.74.

\begin{figure}[h!]
 \includegraphics[trim = 0mm 0mm 0mm 0mm,width=1\columnwidth,clip=true]{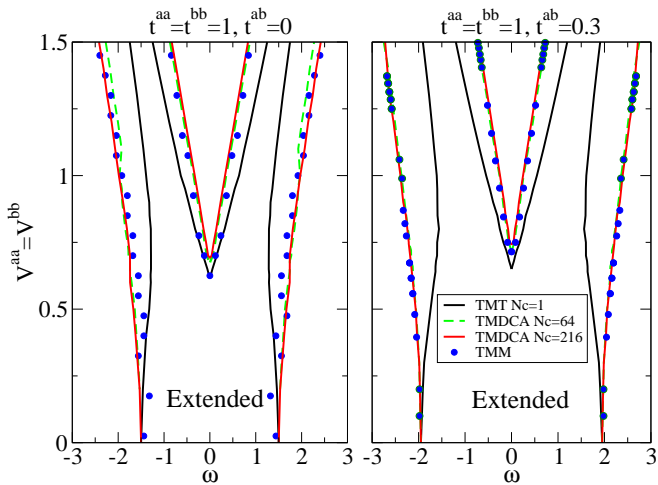}
 \caption{Disorder-energy phase diagram for vanishing $t^{ab}$ (left panel) and finite $t^{ab}=0.3$ (right panel). 
 We compare the mobility edge obtained from the TMT ($N_c=1$), TMDCA ($N_c=64$ and $216$) and TMM. Parameters for the 
 TMM data are given in the text (the TMM data for $t_{ab}=0.0$ is reproduced from ~\cite{PhysRevB.90.094208}).  
 A finite $t^{ab}$ increases the critical disorder strength, indicating that $t^{ab}$ results in a delocalizing 
 effect. The single site TMT overestimates the localized region.}
 \label{fig:mobility_edge}
\end{figure}

To study the effect of inter-band hopping $t^{ab}$, we calculate the disorder-energy phase diagram 
for the case with vanishing $t^{ab}$ and finite $t^{ab}=0.3$ in Fig.~\ref{fig:mobility_edge}.  The
mobility edge 
is determined by the energy where the TDOS vanishes.  By comparing the left and right panels, we can 
see that introducing a finite $t^{ab}$ makes the system more difficult to localize, causing an upward 
shift of the mobility edge.  The single site TMT overestimates the localized region compared to 
finite cluster results. We also compare our results with those from the TMM ~\cite{Markos-review-2006,MacKinnonKramer1983,Kramer2010} 
to check the accuracy of the mobility edge calculated from TMDCA.  For the TMM, the Schr\"{o}dinger equation 
is written in terms of wavefunction amplitudes for adjacent layers in a quasi-one dimensional system, and the
correlation (localization) length is computed by accumulating the Lyapunov exponents of successive 
transfer matrix multiplications that describe the propagation through the system.  All TMM data is 
for a 3d system of length $L = 10^6$ and the Kramer-MacKinnon scaling parameter $\Lambda(V,M)$ is 
computed for a given disorder strength $V$ and ``bar'' width $M$.  The transfer matrix is a $2Ml_b \times 
2Ml_b$ matrix.  The system widths used were $M=[4-12]$.  The critical 
point is found by identifying the crossing of the $\Lambda(M) {\rm vs.} V$ curves for different system sizes. 
The transfer matrix product is reorthogonalized after every five multiplications.
%Note to me, add TMM description here - Conrad

\begin{figure}[h!]
 \includegraphics[trim = 0mm 0mm 0mm 0mm,width=1\columnwidth,clip=true]{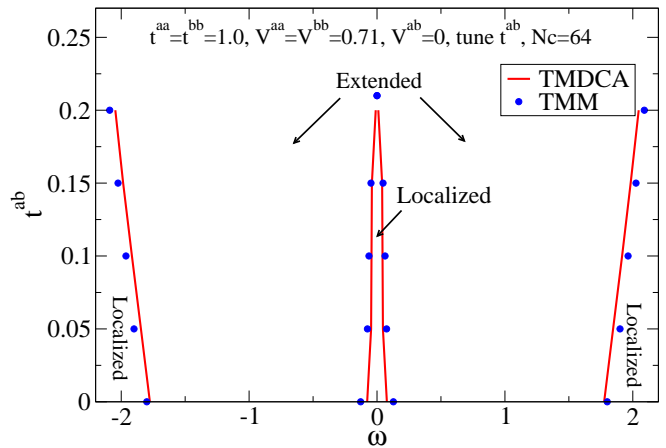}
 \caption{Evolution of the mobility edge as $t^{ab}$ increases, while $V^{aa}$ and $V^{bb}$ are fixed. 
The results are calculated for $N_c=64$. A dome-like shape shows up around the band center, signaling the 
closing of the TDOS gap.  Parameters for the TMM data are given in the text.}
 \label{fig:pd_tune_tab}
\end{figure}

To see the effect of inter-band hopping more directly, we now consider increasing 
$t^{ab}$ while keeping the disorder strength fixed ($V^{aa}=V^{bb}=0.71$), and study the
evolution of the mobility edge (Fig.~\ref{fig:pd_tune_tab}). The localized region around the 
band center starts to shrink as $t^{ab}$ is increased, leading to a small dome-like 
shape with the top located at $t^{ab}=0.2$. This 
shows that increasing $t^{ab}$ delocalizes the system which is reasonable 
since increasing $t^{ab}$ effectively increases the bare bandwidth.

\begin{figure}[h!]
 \includegraphics[trim = 0mm 0mm 0mm 0mm,width=1\columnwidth,clip=true]{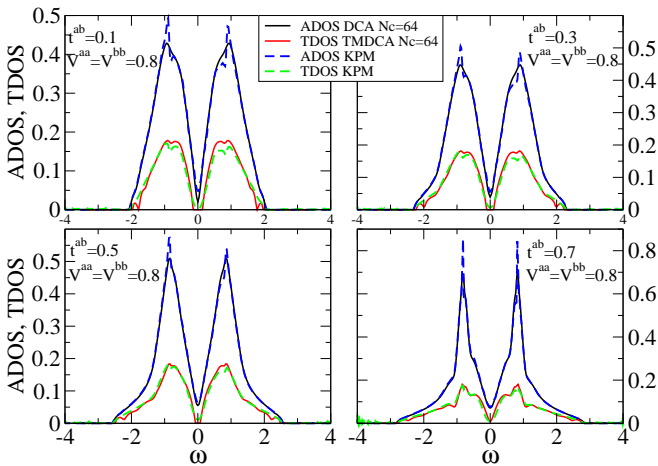}
 \caption{Comparison of ADOS and TDOS calculated with DCA, TMDCA and KPM with fixed disorder strength 
 $V^{aa}=V^{bb}=0.8$ and various values of inter-band hopping $t^{ab}$. The KPM uses 2048 moments on a cubic lattice of
 size $48^3$ and 200 independent realizations generated with 32 sites randomly sampled from each realization.}
 \label{fig:kpm}
\end{figure}

To further benchmark our algorithms, we plot the ADOS and TDOS calculated with two-band DCA and TMDCA together 
with those calculated by the KPM \cite{Schubert, KPM_review_2006, PhysRevB.71.045126, PhysRevB.77.245130} 
(Fig.~\ref{fig:kpm}).  In the KPM analysis, the LDOS is expanded by a series of Chebyshev polynomials, so that 
the ADOS and TDOS can be evaluated. The details for the implementation of KPM are well discussed in 
Ref.~\onlinecite{KPM_review_2006} and the parameters used in the KPM calculations are listed in the caption of 
Fig.~\ref{fig:kpm}. The Jackson kernel is used in the calculations \cite{KPM_review_2006}. As shown in the plots, the results from the generalized DCA and TMDCA match 
nicely with those calculated from the KPM. 

The excellent agreement of the TMDCA results with those from more conventional numerical methods, like
KPM and TMM, suggest that the method may be used for the accurate study of real materials.

\subsection{Application to K$_{\lowercase{y}}$F\lowercase{e}$_{2-\lowercase{x}}$S\lowercase{e}$_2$}\label{sec:KFeSe}

Next, we demonstrate the method with a case study of Fe vacancies in the Fe-based superconductor K$_x$Fe$_{2-y}$Se$_2$, 
which has been studied intensely because of its peculiar electronic and structural properties.
Early on it was found that there is a strong $\sqrt{5}\times\sqrt{5}$ ordering of Fe vacancies \cite{wbao}.
Later it was discovered that this material also contains a second phase\cite{ricci,zwang}.
It is commonly speculated that the second phase is the one that hosts the superconducting state and the phase with 
the $\sqrt{5}\times\sqrt{5}$ vacancy ordering is an antiferromagnetic (AFM) insulator.
Recent measurements of the local chemical composition \cite{landsgesell,hhwen278} 
have determined that the second phase also contains a large concentration of Fe vacancies (up to 12.5\%).
However, these Fe vacancies are not well ordered since no strong reconstruction of the Fermi surface \cite{chlin,tberlijn,ccao} 
was observed by angle-resolved photoelectron spectroscopy (ARPES) experiments~\cite{PhysRevX.1.021020,Zhang2011}.

Interestingly, with such a disordered structure, this material hosts a relatively high superconducting transition 
temperature of 31 $K$ at ambient pressure~\cite{PhysRevB.82.180520}. 
It was the first Fe-based superconductor that was shown from ARPES~\cite{PhysRevX.1.021020,Zhang2011} 
to have a Fermi surface with electron pockets only and no hole pockets, 
apparently disfavoring the widely discussed $S^{\pm}$ pairing symmetry~\cite{mazinSpm} in the Fe-based superconductors.
K$_x$Fe$_{2-y}$Se$_2$ is also the only Fe-based superconductor whose parent compound (with perfectly ordered Fe vacancy) 
is an AFM insulator~\cite{0295-5075-94-2-27009} rather than a AFM bad metal. 
Furthermore from neutron scattering~\cite{wbao}, it has been observed that the anti-ferromagnetism has a novel block type 
structure with a record high Neel temperature of $T_N = 559 K$ and magnetic moment of 3.31$\mu_B$/Fe.
Such a special magnetic structure is obviously not driven from the nesting of the simple Fermi surface, 
but requires the interplay between local moments and itinerant carriers present in the normal state~\cite{yin2010,tam2015}.

Given that Fe vacancies are about the strongest possible type of disorder that can exist in Fe-based superconductors 
and given that the Fe-based superconductors are quasi two-dimensional materials, it is natural to speculate how close 
the second phase is to an Anderson insulator.
If it is indeed close, this would have interesting implications for the strong correlation physics and the non-conventional 
superconductivity in these compounds.

\begin{figure}[h!]
 \includegraphics[trim = 0mm 0mm 0mm 0mm,width=1\columnwidth,clip=true]{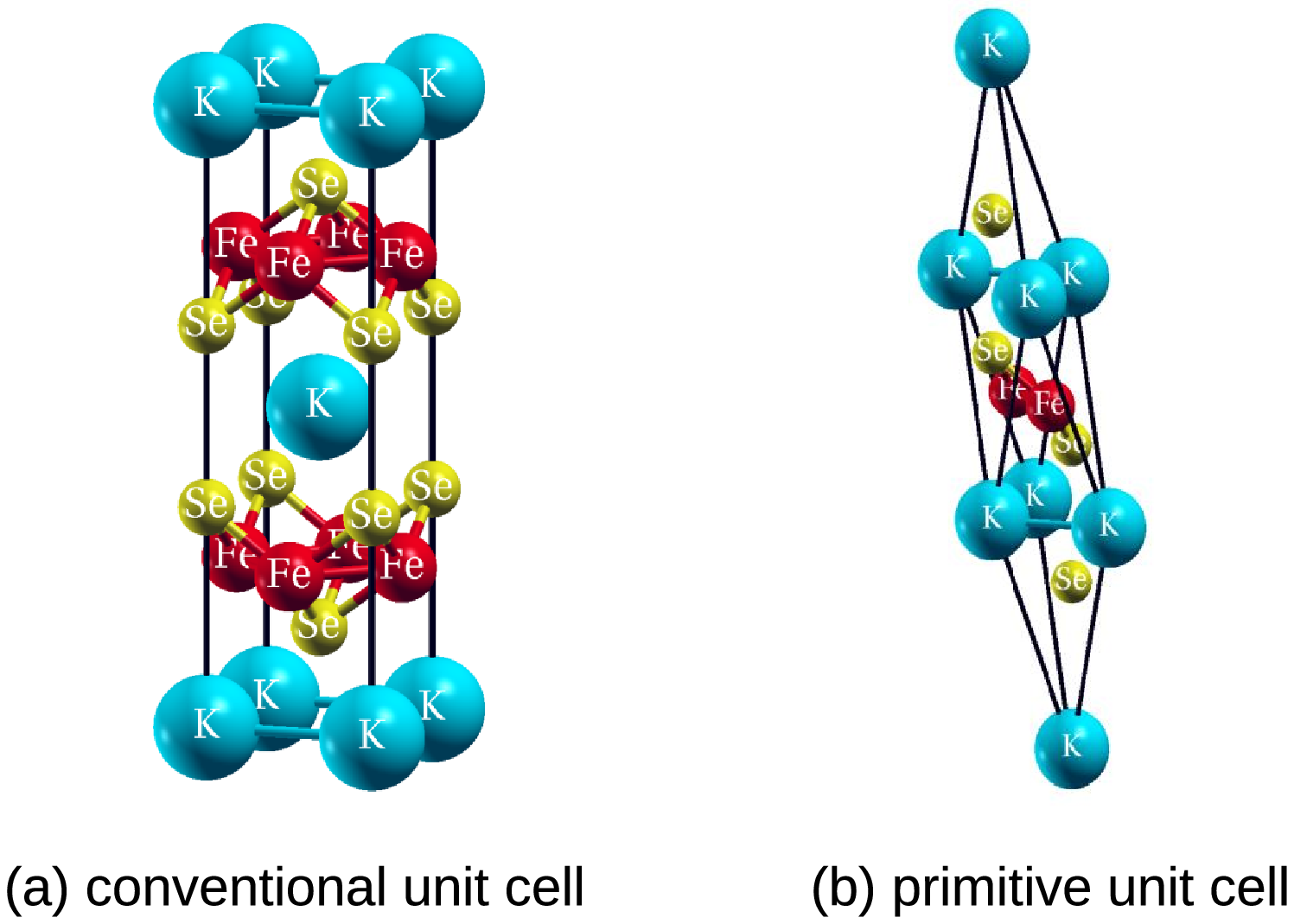}
 \caption{Crystal structure of KFe$_2$Se$_2$.}
 \label{fig:KFeSe}
\end{figure}

To investigate the possibility of Anderson localization in the second phase of K$_x$Fe$_{2-y}$Se$_2$ we will employ 
TMDCA on a realistic first principles model. To this end we use Density Functional Theory (DFT) in combination
with the projected Wannier function technique \cite{kuwannier} to extract the low energy effective 
Hamiltonian of the Fe-$d$ degrees of freedom. Specifically we applied the WIEN2K~\cite{wien2k} 
implementation of the full potential linearized augmented plane wave method in the local density approximation.
The k-point mesh was taken to be $10\times10\times10$ and the basis set size was determined by RKmax=7.
The lattice parameters of the primitive unit cell (c.f. Fig.~\ref{fig:KFeSe}(b)) are taken from Ref.~\onlinecite{wbao}.
The subsequent Wannier transformation was defined by projecting the Fe-d characters on the low energy bands within the 
interval [-3,2]~eV.
For numerical convenience, we use the conventional unit cell shown in Fig.~\ref{fig:KFeSe}(a) which contains 4 Fe atoms. 
Since there are 5 $d$ orbitals per Fe atom, we are dealing with a 20-band problem. 
To simulate the effect of Fe vacancies we add a local binary disorder with strength $V$ and Fe vacancy concentration $c_a$:
\begin{equation}
 P(V_{i})=c_a \delta(V_{i}-V)+(1-c_a) \delta(V_{i}).
\end{equation}
We set the disorder strength to be $V=20eV$, much larger than the Fe-d bandwidth, such that it effectively removes 
the corresponding Fe-$d$ orbitals from the low energy Hilbert space.
This will capture the most dominant effect of the Fe vacancies. 
The Fe concentration is taken to be $c_a=12.5\%$, which is the maximum value found in the experiments.

\begin{figure}[h!]
 \includegraphics[trim = 0mm 0mm 0mm 0mm,width=1\columnwidth,clip=true]{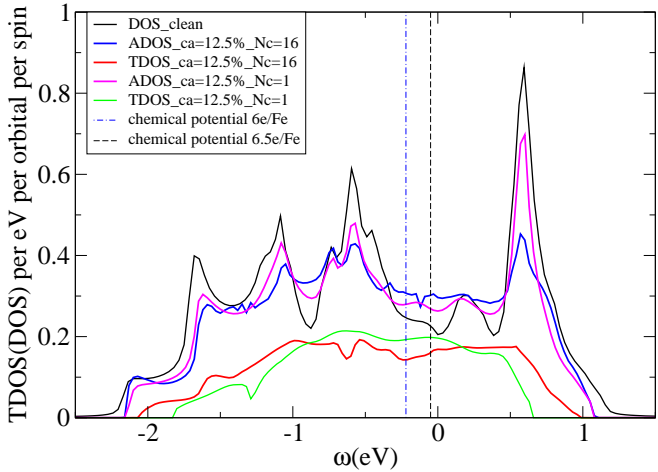}
 \caption{The average and typical density of states of KFe$_2$Se$_2$ with 12.5\% Fe vacancy concentration calculated by 
 multiband DCA and TMDCA with cluster size $N_c=1$ and $N_c=16$, compared with the average density of states of the clean (no vacancy) KFe$_2$Se$_2$. 
 }
 \label{fig:KFeSe_tdos}
\end{figure}

Fig.~\ref{fig:KFeSe_tdos} presents the ADOS and TDOS, obtained from our multiband TMDCA for which we considered 
two cluster sizes $N_c=1$ and $N_c=2\sqrt{2}\times2\sqrt{2}\times2=16$.
Consistent with the model calculations presented in the previous sections, we find that 
the TMT ($N_c=1$) tends to overestimate the localization effects compared to TMDCA results ($N_c=16$). 
While the TMT shows localized states within [0.6,1.1]~eV, the TMDCA for $N_c=16$ finds localized states in 
the much smaller energy region [1.0,1.1]~eV instead.
Apparently a concentration of $c_a=12.5\%$ is still too small to cause any significant localization effects 
despite the strong impurity potentials of the Fe vacancies and the material being quasi-two dimensional.
To determine the chemical potential we consider two fillings.
The first filling of 6.0 electrons per Fe corresponds to the reported K$_2$Fe$_7$Se$_8$ phase~\cite{hhwen278}.
Since strong electron doping has been found in ARPES experiments~\cite{PhysRevX.1.021020,Zhang2011}, we also consider 
a filling of 6.5 electrons per Fe. The latter would correspond to the extreme case of no vacancies.
Clearly for both fillings the chemical potential remains energetically very far from the 
mobility edge, and thus far from Anderson insulating.

\section{Conclusion}\label{sec:conclusion}
We extend the single-band TMDCA to multiband systems and study electron localization for a 
two-band model with various hopping and disorder parameters.  We benchmark our method by
comparing our results with those from other numerical methods (TMM and KPM) and find 
good agreement. We find that the inter-band hopping leads to a delocalization effect, since 
it gradually closes the $\omega=0$ disorder induced gap on the TDOS.
A direct application of our 
extended TMDCA could be done for disordered systems with strong spin-orbital coupling. Combined 
with electronic structure calculations, our method can be used to study the electron localization 
phenomenon in real materials. To show this, we apply this approach to the iron selenide 
superconductors K$_x$Fe$_{2-y}$Se$_2$ with Fe vacancies. By calculating the TDOS around the 
chemical potential, we conclude that the insulating behavior of its normal state is unlikely 
due to Anderson localization.  This method also has the ability to include 
interactions~\cite{2015arXiv150300025E}, and future work will involve real material
calculations that fully treat both disorder and interactions.

\textit{Acknowledgments}--
This work is supported in part by the National Science Foundation 
under the NSF EPSCoR Cooperative Agreement No. EPS-1003897 with additional 
support from the Louisiana Board of Regents (YZ, HT, CM, CE, KT, JM, and MJ).
Work by TB was performed at the Center for Nanophase Materials Sciences, a DOE Office 
of Science user facility. This manuscript has been authored by UT-Battelle, LLC
under Contract No. DE-AC05-00OR22725 with the U.S.
Department of Energy. WK acknowledges support from U.S. Department of Energy, Office of Basic 
Energy Science, Contract No. DEAC02-98CH10886. The United States Government retains
and the publisher, by accepting the article for publication,
acknowledges that the United States Government retains a
non-exclusive, paid-up, irrevocable, world-wide license to
publish or reproduce the published form of this manuscript, or
allow others to do so, for United States Government purposes.
The Department of Energy will provide public access to these
results of federally sponsored research in accordance with the
DOE Public Access Plan (http://energy.gov/downloads/doepublic-
access-plan). This work used the high performance computational 
resources provided by the Louisiana Optical Network Initiative (http://www.loni.org),
and HPC@LSU computing. 

\appendix

\section{The order parameter defined in Eq.~\ref{eqn:op}}
\label{appendixA}
We know the system is localized if the distribution of the total LDOS is critical, having 
a probability 
distribution $p(\rho_i^{aa}+\rho_i^{bb})$ which is highly skewed with a typical value close to zero.
So if we can show that this is true if and only if both $\rho_i^{aa}$ and $\rho_i^{bb}$ are critical, 
then 
the critical behavior is basis independent and we can choose any particular basis 
and use the order parameter defined by Eq.~\ref{eqn:op} to study the 
localization transition.

To show this is true, we consider two probability distribution 
functions $p_1(x_1)$ and $p_2(x_2)$.
The probability distribution function for $X=x_1+x_2$ is
\begin{equation}
 P(X)=%\int_{0}^{\infty}p_{1}(x)p_{2}(X-x)dx=
\int_{0}^{X}p_{1}(x)p_{2}(X-x)dx,
\end{equation}
and we want to show $P(X)$ is critical if and only if both $p_1(x_1)$ and $p_2(x_2)$ are critical.

\subsection{Sufficiency}
If both $p_1(x)$ and $p_2(x)$ are critical, then both $p_1(x)$ and $p_2(x)$ are dominated by the 
region $0<x<\delta$ where $\delta\rightarrow0^{+}$.
The contribution to the integral in $P(X)$ mainly comes from the 
region $0<x<\delta$ and $0<X-x<\delta$ which is $max(X-\delta,0)<x<min(\delta,X)$. 
Since $\delta$ is infinitesimal, we can assume $X>\delta$, and then we have $X-\delta<x<\delta$.
To maximize $P(X)$, we want this region to be as big as possible, so
we want $\delta-(X-\delta)=2\delta-X$ to be as big as possible which 
means $X$ must be smaller than $2\delta\rightarrow0^{+}$.  Thus, $P(X)$ is also critical with the 
typical value around $2\delta$ which is infinitesimal. 

\subsection{Necessity}
We now consider the case where one of the distributions is not critical.
Without loss of generality, we assume $p_2(x)$ is not critical and is peaked at some finite value $x_0$.
We calculate
\begin{widetext}
\begin{equation}
\begin{split}
%P(x_0)-P(\delta)&=\int_0^{x_0} p_1(x) p_2(x_0-x) dx \\ 
%&- \int_0^{\delta} p_1(x) p_2(\delta-x) dx\\
%&=\int_0^{\delta} p_1(x) [p_2(x_0-x) - p_2(\delta-x)] dx\\
%&+ \int_{\delta}^{x_0} p_1(x) p_2(x_0-x) dx
P(x_0)-P(\delta)&=\int_0^{x_0} p_1(x) p_2(x_0-x) dx  - \int_0^{\delta} p_1(x) p_2(\delta-x) dx\\
&=\int_0^{\delta} p_1(x) [p_2(x_0-x) - p_2(\delta-x)] dx
+ \int_{\delta}^{x_0} p_1(x) p_2(x_0-x) dx.
\end{split}
\end{equation}
\end{widetext}
The first term is positive since $p_2(x)$ is peaked around $x_0$ and $\delta\ll x_0$.  The second term is positive 
obviously, so $P(x_0)>P(\delta)$.  Therefore, $P(X)$ is not critical. 

In this way we argue that $P(X)$ is critical if and only if both $p_1(x_1)$ and $p_2(x_2)$ are critical. 
In other words, when the localized states hybridize
with extended states, only extended states remain which is exactly Mott's insight about the mobility edge~\cite{Mott}.
The generalization to the multiple band case is trivial.

\bibliography{TMDCA_2band}
 
\end{document}